\documentclass[conference]{IEEEtran}
\usepackage[english]{babel}
\usepackage{amsmath}
\usepackage{amssymb}
\usepackage{amsbsy}
\usepackage{amsfonts}

\usepackage{color}
\usepackage{subfig}
\usepackage{graphicx}

\usepackage{booktabs}
\usepackage{colortbl}

\usepackage{enumerate}

\usepackage{listings}
\usepackage{caption}
\DeclareCaptionFont{white}{\color{white}}
\DeclareCaptionFormat{listing}{\colorbox[cmyk]{0.43, 0.35, 0.35,0.01}{\parbox{\textwidth}{\centering #1#2#3}}}
\newcommand{\chol}{{\sc CLAK-Chol}}
\newcommand{\eig}{{\sc CLAK-Eig}}

\DeclareCaptionType{copyrightbox}

\begin{document}

\title{High-throughput genome-wide association analysis for single and multiple phenotypes}
\author{\IEEEauthorblockN{Diego Fabregat-Traver}
\IEEEauthorblockA{
Aachen Institute for Advanced Study\\
in Computational Engineering Science\\
RWTH Aachen, Aachen, 52062, Germany\\
Email: fabregat@aices.rwth-aachen.de}
\and
\IEEEauthorblockN{Yurii S. Aulchenko}
\IEEEauthorblockA{Institute of Cytology and Genetics\\
SD RAS, Novosibirsk, 630090, Russia\\
Email: yurii.aulchenko@gmail.com}
\and
\IEEEauthorblockN{Paolo Bientinesi}
\IEEEauthorblockA{Aachen Institute for Advanced Study\\
in Computational Engineering Science\\
RWTH Aachen, Aachen, 52062, Germany\\
Email: pauldj@aices.rwth-aachen.de}}

\maketitle

\begin{abstract}
The variance component tests used in genome-wide association studies of thousands of individuals become computationally exhaustive when multiple traits are analysed in the context of omics studies. We introduce two high-throughput algorithms ---\chol{} and \eig{}--- for single and multiple phenotype genome-wide association studies (GWAS). The algorithms, generated with the help of an expert system, reduce the computational complexity to the point that thousands of traits can be analyzed for association with millions of polymorphisms in a course of days on a standard workstation. By taking advantage of problem specific knowledge, \chol{} and \eig{} significantly outperform the current state-of-the-art tools in both single and multiple trait analysis.
\end{abstract}

Current biomedical research is experiencing a large boost in the amount of data
generated. Individual genomes are being characterized at increased level of
details using single nucleotide polymorphism (SNP) arrays, and, more recently,
exome and whole-genome re-sequencing. At the same time, technologies for high-throughput
characterization of tens of thousands of molecular “omics” phenotypes in thousands of people are becoming increasingly
affordable~\cite{Goring2007}~\cite{10.1371/journal.pgen.1000672,10.1371/journal.pgen.1002490}~\cite{10.1371/journal.pgen.1001256}.

Genome-Wide Association Studies (GWAS) is a powerful tool for identifying loci
involved in the control of complex traits~\cite{Hindorff-2009}. In such studies, thousands of
individuals are measured for the trait of interest and their genomes are
characterized by re-sequencing or by genotyping of millions of genetic markers
using SNP arrays. The association between genetic markers and a phenotype of
interest is studied, with significant association highlighting the genomic
regions harboring functional variants involved in the control of the trait.

Even in carefully designed population-based studies, some degree of relatedness
and population stratification is expected. One of the most flexible and powerful
methods of accounting for such substructure is the
Variance Components (VC) approach based on linear mixed models~\cite{Yu-Pressoir-2006,Astle-Balding-2009} ({\bf Supplementary Note, Section 1}). Recent FaST-LMM~\cite{Lippert2011} implementation of the VC model can complete
genome-wide scan of association between 36 millions of genetic markers and a
phenotype in 1,000 individuals in about 4 hours using a multicore(12) processor.
While these results are impressive and
represent a breakthrough compared to older algorithms and implementations, such
computational throughput is still prohibitively low for analysis of “omics” data.
For example, for transcriptome (30,000 phenotypes), the estimated time to complete the analysis would be
approximately 13 years. Currently, such
amount of computations would only be tractable if large computing
facilities with many thousands of cores are used.

\begin{figure*}
    \centering
    \includegraphics[scale=.65]{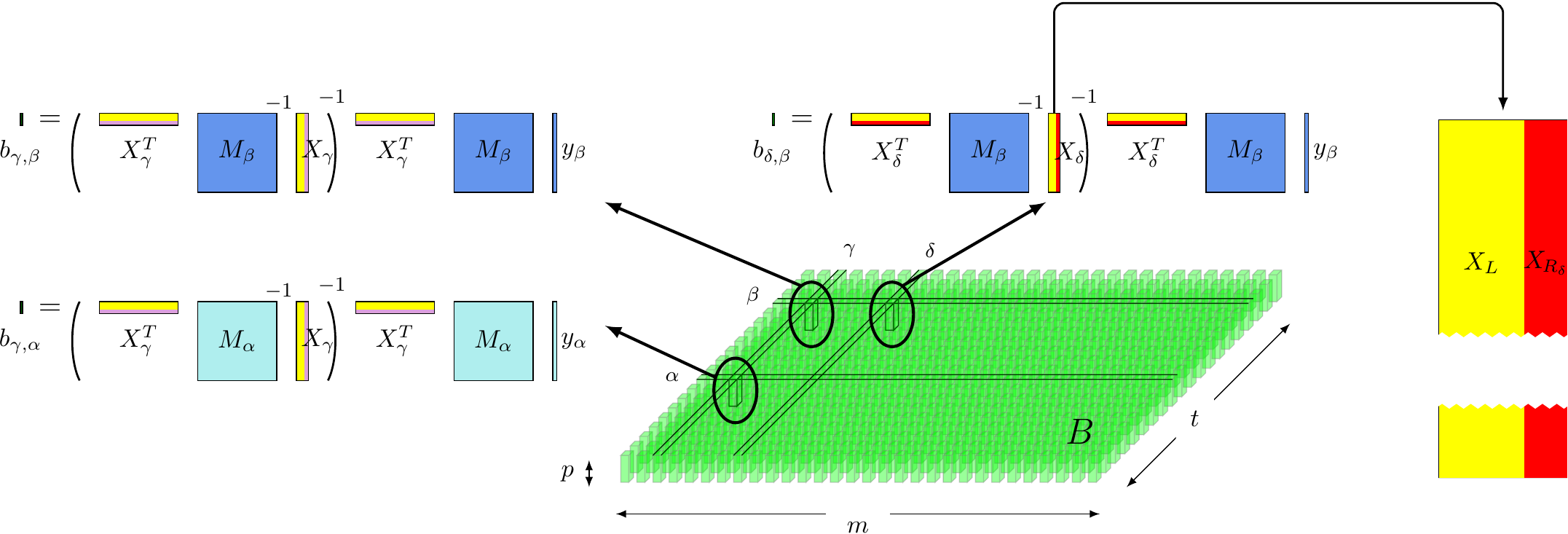}
\caption{Interpretation of GWAS as a 2-dimensional sequence of generalized least-squares problems ($ b := (X^T { M }^{-1} X)^{-1} X^T { M }^{-1} y$).
GWAS with multiple phenotypes requires the solution of $m \times t$ correlated
GLS problems, originating a three-dimensional object $B$ of size $m \times t \times p$.  
Along the $t$ direction, the covariance matrix $M$ and the phenotype
$y$ vary, while the design matrix $X$ does not; conversely, in the $m$
direction, $M$ and $y$ are fixed while $X$ varies. Specifically,
$X$ can be viewed as consisting of two parts, $X_L$ and $X_R$, where
the former is constant across the entire grid and the latter changes along $m$.
The figure also captures GWAS with single phenotype, in which case the dimension
$t$ reduces to 1.}
\label{fig:ProblemDescription}
\end{figure*}

In this work, we consider the problem of genome-wide association analysis using
the mixed models, with special emphasis on analysis of thousands of phenotypes
(Fig.~\ref{fig:ProblemDescription}). We demonstrate how utilization of problem-specific
knowledge, coupled with innovative techniques of automatic generation of
linear algebra algorithms and hardware-tailored implementation, allows reducing the computation time from years to days. For the aforementioned analysis of transcriptome, our algorithm takes 26 days on a single node, or equivalently, 20 hours on a standard 32-node cluster.

In the context of linear algebra computations, we recently developed a symbolic system, CLAK, that closely mirrors the reasoning of a human expert for the generation and analysis of algorithms~\cite{CLAK}. The idea is to first decompose a target linear algebra operation in terms of library-supported kernels, and then apply optimizations aimed at reducing redundant calculations. Since the decomposition is not unique, CLAK returns not one but a family of algorithms, together with their corresponding cost estimates.

Figure~\ref{fig:ProblemDescription} illustrates how multi-trait analysis consists of $t$ separate single-trait analyses, each of which, in turn, consists of $m$ generalized least-squares (GLS) problems. The key to fast algorithms is the realization that such problems are correlated, both in the $m$ and $t$ directions; a naive approach that only aims at optimizing one GLS in isolation will never be competitive with methods that tackle sequences as a whole.
With the help of CLAK, we generated more than 20 algorithms for performing genome-wide association analysis. Based on the cost estimates, we assessed the potential of these algorithms for single-trait and multiple-trait studies. Interestingly, for the two scenarios, the best theoretical performance was attained by two different algorithms, \chol{} and \eig{}, respectively ({\bf Supplementary Note, Sections 2 and 3}).

Both algorithms incorporate a number of techniques for increasing the computational efficiency and save intermediate results across adjacent problems. In \chol{}, differently than in the other existing methods (such as EMMAX~\cite{Kang-2010} and FaST-LMM), the positive-definiteness of the covariance matrix is exploited to utilize a faster factorization; moreover, the computation is rearranged to fully benefit from the potential of the underlying optimized kernels.
In \eig{} instead, redundant computation is avoided by exploiting the fact that the relationship (kinship) matrix is constant even across different traits, and only the heritability and total variance of the trait change. When compared with the current state-of-the-art implementations such as FaST-LMM, both our algorithms achieve a lower computational complexity (for a comparison, see {\bf Supplementary Table~1}).

\begin{figure*}
    \centering
    \subfloat[]{\includegraphics[scale=0.55]{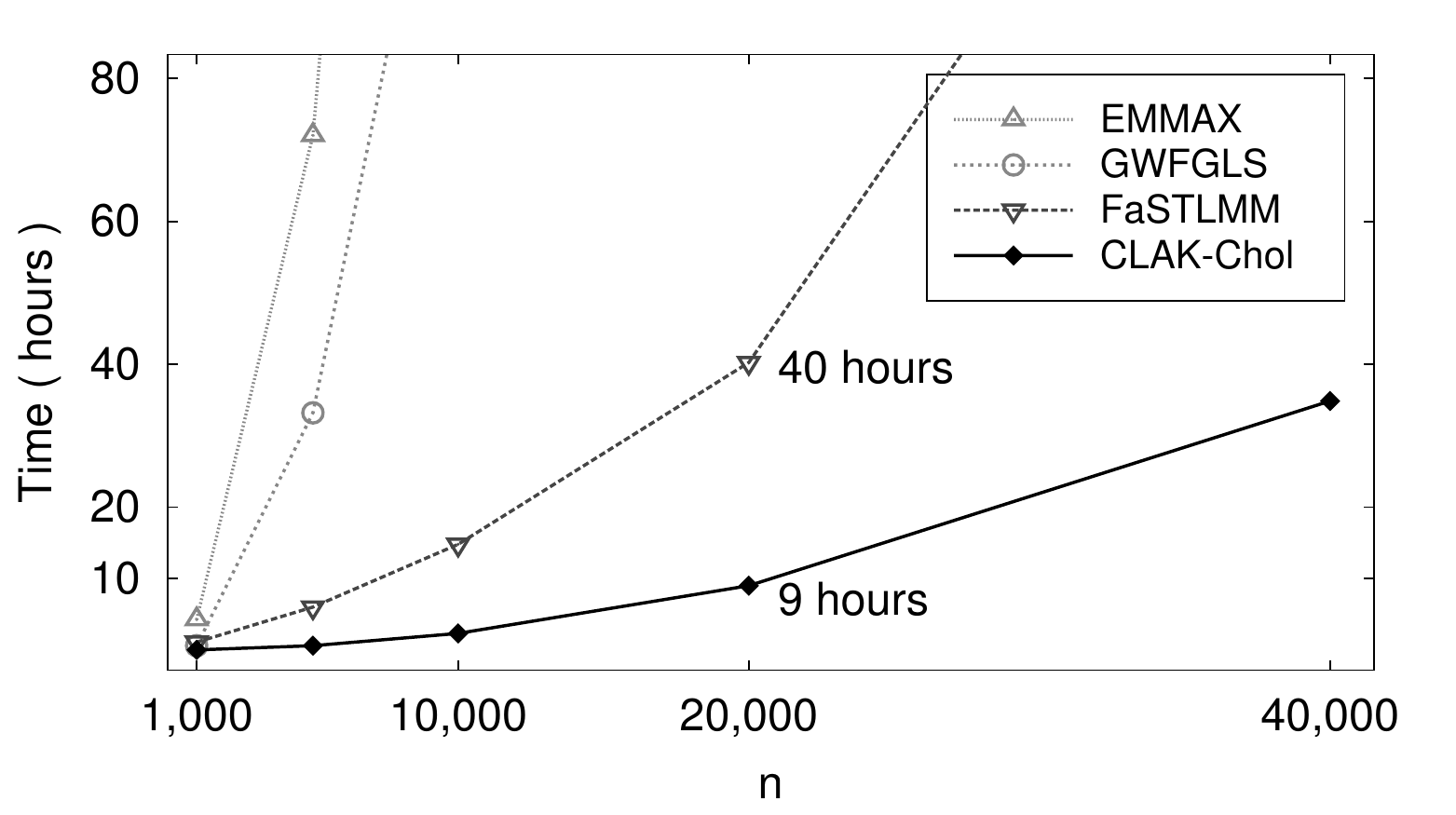}} \quad
    \subfloat[]{\includegraphics[scale=0.55]{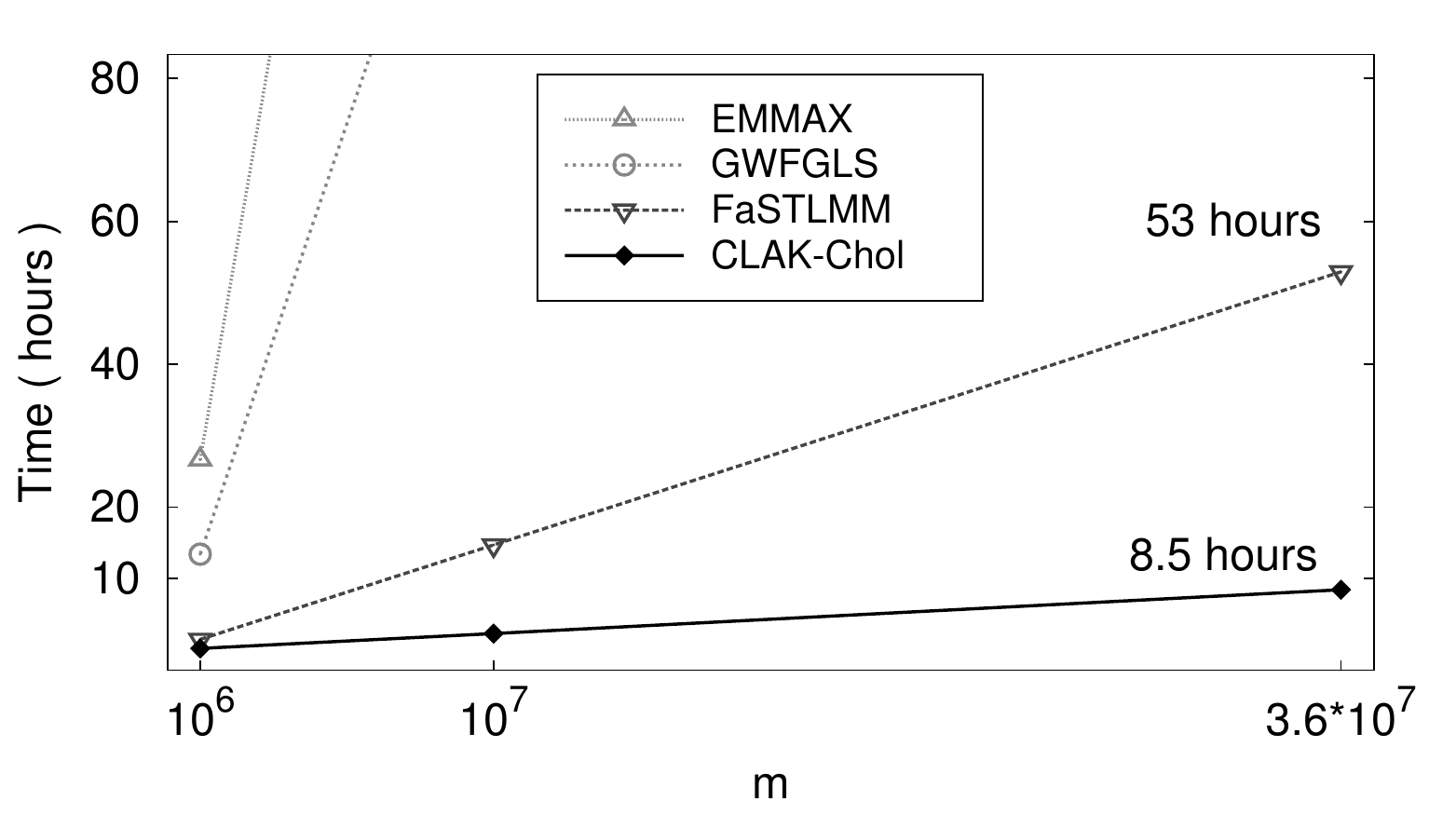}} \\
    \subfloat[]{\includegraphics[scale=0.55]{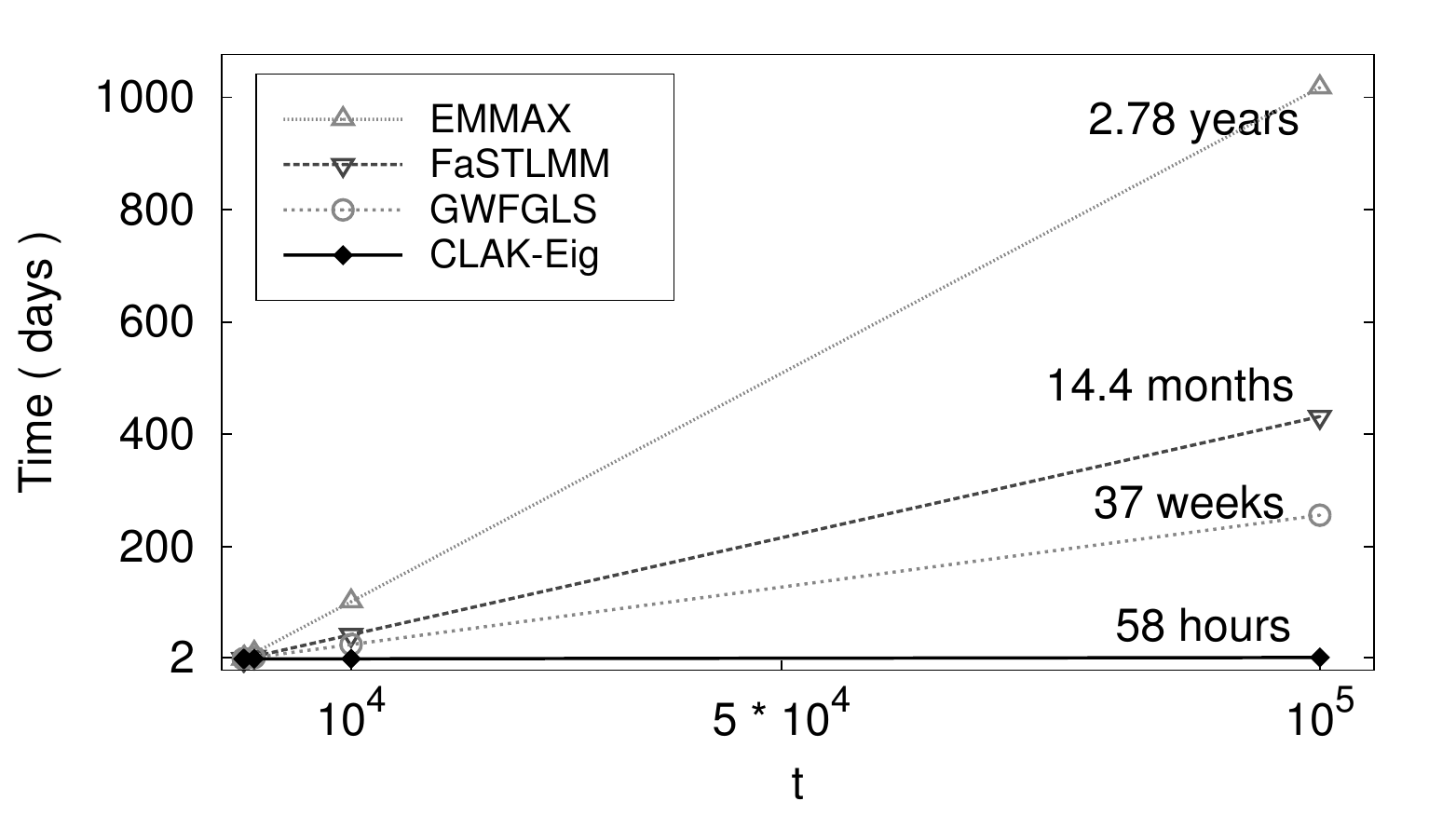}} \quad
    \subfloat[]{\includegraphics[scale=0.55]{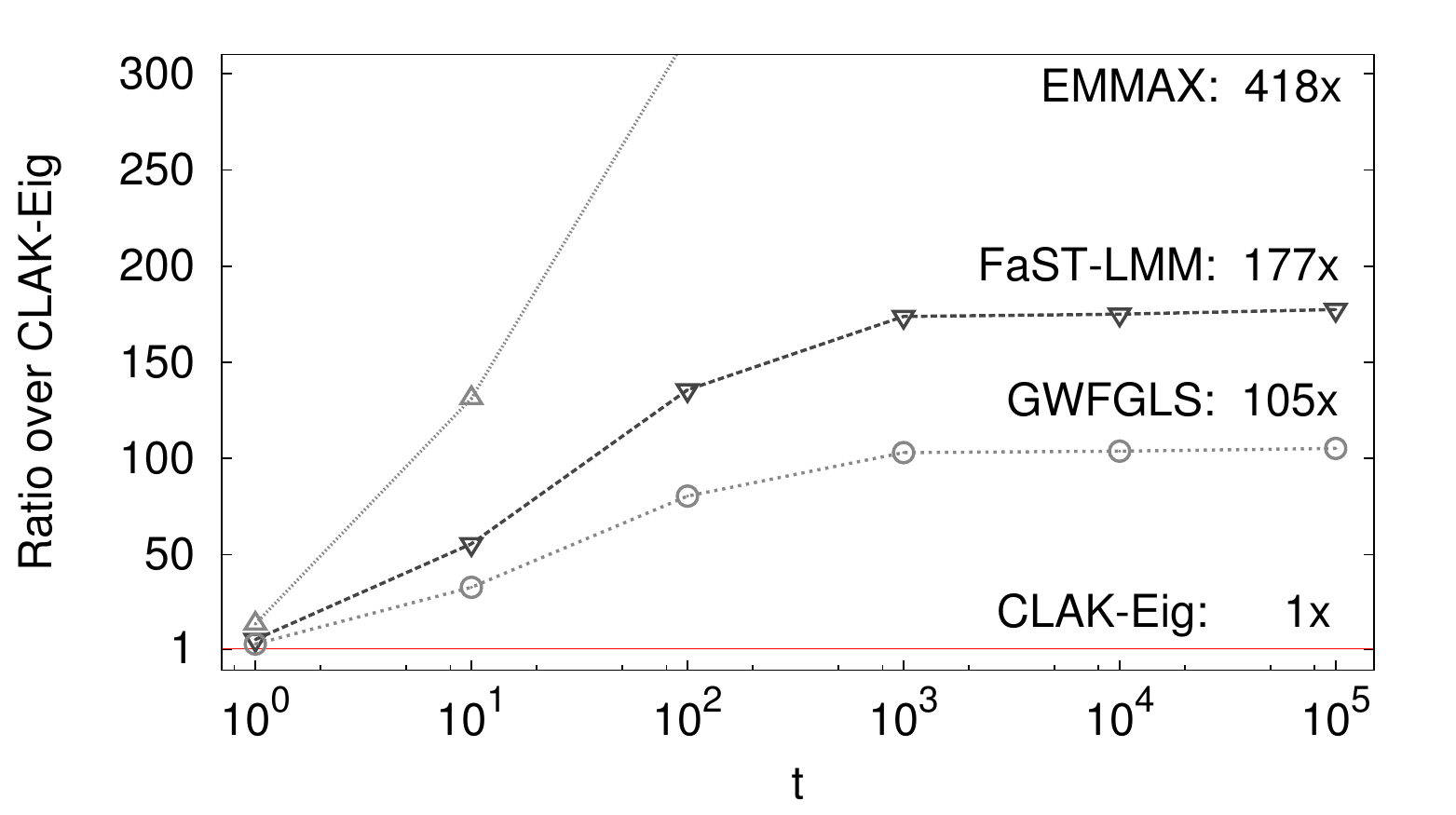}}
  \caption{Timing comparison.
  Panels (a) and (b) include timings for EMMAX, GWFGLS, FaST-LMM, and
\chol{}, relative to single trait analysis;
  (c) and (d) present a comparison of EMMAX, GWFGLS, FaST-LMM, and
\eig{} in the case of multiple traits.
  In (a), the number of SNPs is fixed to $m = 10{,}000{,}000$ and the
sample size $n$ ranges from $1{,}000$ to $40{,}000$.
  In panel (b), the sample size is fixed to $n=10{,}000$ and the number of
SNPs $m$ ranges between $10^6$ and $3.6 \times 10^7$.
  In (c) and (d), $n=1{,}000$, $m=10^6$ and $t$ ranges from 1 to
$100{,}000$.
}
\label{fig:results}
\end{figure*}

The size of the data sets involved in genome-wide association studies is considerably larger than the memory capacity of current processors; input and output data can only be stored in disk devices. Since the penalty for accessing a piece of data residing on disk is enormous --several order of magnitude greater than the cost for performing one arithmetic operation-- it is imperative to efficiently handle the data. From our experiments, in fact, we observed that the time spent on data movement adds about 30\% to the time spent on arithmetic calculations. Instead, through asynchronous transfers between memory and disk, our algorithms achieve a perfect overlap of computation and data movement. As long as the covariance matrix fits in main memory, and regardless of the size of the data sets --both in terms of SNPs and phenotypes--, the processor never idles waiting for data, thus computing at maximum efficiency.

To demonstrate the practical advantages of \eig{} and \chol{}, we implemented routines and compared their execution time with that of well-established methods: EMMAX, FaST-LMM (two-step approximation), and GWFGLS (implementation of the mmscore method of ProbABEL~\cite{Aulchenko-Struchalin-2010} in the MixABEL-package). In the experiments we considered three different scenarios, varying the sample size, the number of SNPs, and the number of traits, while keeping the other two values constant (Fig.~\ref{fig:results}). A description of the experimental setup is provided in the {\bf Supplementary Note, Section 4}.

In the first scenario (single trait and $10{,}000{,}000$ SNPs), even though all methods exhibit a quadratic behaviour, \chol{} is the only algorithm that completed all tests within 1.5 days. For the largest problem considered (sample size $n = 40{,}000$), the speedup over FaST-LMM is 4.53: 158 hours vs.~35; for $n = 1{,}000$ instead, the speedups over GWFGLS, FaST-LMM and EMMAX are 15, 28 and 106, respectively: 38, 68 and 257 minutes vs. 2.5 minutes.

The second scenario (single trait and sample size of $10{,}000$) shows a linear dependence on the number of genetic markers for all methods. Again, \chol{} attains the best timings, outperforming FaST-LMM, GWFGLS and EMMAX by a factor of 6.3, 56.8 and 112, respectively.

Thanks to \eig{}'s linear complexity with respect to sample size, SNPs, and traits,
the advantage in the analysis of multiple phenotypes (third scenario: sample size of ${1,000}$ and $1{,}000{,}000$ genetic markers) becomes most apparent:
when thousands and more traits are considered, \eig{} outperforms GWFGLS, FaST-LMM and EMMAX by a factor of 105, 177, and 418, respectively, bringing the execution time from several months down to two days.

Establishing the functional roles of genetic variants and finding the biological link between genetic variation and complex phenotypes remain significant challenges in the post-genomic era. Genome-wide association studies are increasingly applied to understand the regulation of human and animal transcriptome~\cite{Goring2007}, metabolome~\cite{10.1371/journal.pgen.1000672,10.1371/journal.pgen.1002490}, glycome~\cite{10.1371/journal.pgen.1001256} and other types of omics data; they are also used to uncover the link between these molecular phenotypes and high-level complex traits, including common diseases~\cite{Cookson2009}. In GWAS, it is well recognized that genetic (sub)structure can act as confounder and lead to false-positive discoveries, unless correctly accounted for; one of the most flexible and powerful methods for such a correction is provided by the mixed models~\cite{Yu-Pressoir-2006}. However, this powerful method comes at a high computational price: Even when using the most advanced methods and tools~\cite{Lippert2011}, it takes a few hours for the GWAS to complete. While this time scale might not seem a big hindrance when only a relatively small number of phenotypes are analyzed (for instance in studies of complex traits and common diseases), the issues become apparent as soon as omics data are considered. The computational time becomes prohibitively large (up to several years), and in order to complete the studies within reasonable time one has to count on supercomputers with thousands of cores. This solution delays the research and increases its cost.

In this work, we clearly demonstrate that use of problem-specific knowledge, coupled with innovative techniques for algorithm generation and with hardware-tailored implementations, leads to both a decrease in computational complexity and an increase in algorithmic efficiency. Specifically, for the analysis of omics data we were able to attain remarkable speed-ups, making it feasible to analyze ---on a single standard multi-core computer, as opposed to supercomputing facilities--- tens and even hundreds of thousands of phenotypes in the course of few days rather than years.

Further optimizations are possible, for instance by exploiting the structure of the kinship matrix. A “compressed MLM” approach was proposed for decreasing the effective sample size of datasets by clustering individuals into groups~\cite{Zhang-2010}; similarly, the fast decaying and possibly sparse structure of the kinship matrix can be exploited to decrease the algorithmic complexity.

We believe that the approach outlined here ---integrating problem-specific
knowledge with automatic algorithm generation and hardware-tailored
implementation---- has many applications in high-throughput analysis of biomedical
data.


\section{Acknowledgements}
The work of YSA was funded by grants from the Russian Foundation of Basic Research (RFBR), the Helmholtz society (RFBR-Helmholtz Joint Research Groups), and the MIMOmics project supported by FP7.

PB and DFT gratefully acknowledge the support received from the Deutsche Forschungsgemeinschaft (German Research Association) through grant GSC 111.

The authors wish to thank the Center for Computing and Communication at RWTH Aachen for the computing resources.


\end{document}


\title{Supplementary Notes}
\author{Diego Fabregat, Yurii Aulchenko, Paolo Bientinesi}

\maketitle


\section{Mixed Models for GWAS}
\label{MixModels}

The Variance Components model for a quantitative trait can be formulated as
$$ Y = X \beta + R, $$
where $Y$ is the vector containing the phenotypes for $n$ individuals, $X$ is the design matrix, and $\beta$ and $R$ are the vectors of fixed and random effects, respectively. The partitioning $X = [1 | L | g ]$ indicates that the design matrix is composed of three parts: $1$ denotes a column-vector (corresponding to the intercept) containing ones, $L$ is an $n \times p$ matrix corresponding to fixed covariates such as age and sex, and $g$ typically consists of a single column-vector containing genotypes. The vector of random effect $R$ is assumed to be distributed as a Multivariate Normal with mean zero and variance-covariance matrix $M = \sigma^2 \cdot ( h^2 \Phi + (1-h^2) I )$; here, $\sigma^2$ is the total variance of the trait, $h^2$ (in the range [0, 1]) is the heritability coefficient --which defines the strength of the correlation between phenotypes of relatives--, $I$ is the identity matrix, and $\Phi$ is the matrix containing the relationship coefficients for all pairs of studied individuals. The relationship coefficient is defined as the proportion of the genome “identical-by-descent” that a pair of individuals share; for example, in the case of identical twins (they have the same DNA) the relationship coefficient is 1, while since a parent transmits half of its genome to the offspring, the relationship coefficient for a parent with its offspring is 0.5. The relationship matrix $\Phi$ can be estimated from the pedigree or from the genomic data~\cite{Astle-Balding-2009}. In GWAS, the quantity of interest is the effect of the genotype, that is, the element(s) of $\beta$ corresponding to $g$. Technically, a GWAS with mixed model consists of traversing all measured polymorphic sites in the genome, substituting the corresponding $g$ into $X$, and fitting the above model; the result is millions of estimates of genetic effect together with their $p$-values.

One of the most used mixed model-based approaches used in GWAS relies on a two-step analysis methodology~\cite{Chen-Abecasis-2009,Aulchenko-Struchalin-2010,Kang-2010,Zhang-2010,Lippert2011}. In the first step, the reduced model (with $X = [1 | L ]$) is fit to the data, thus obtaining the estimates $\{ \hat{\sigma^2}, \hat{h^2} \}$; the variance-covariance matrix corresponding to such estimates is denoted by $ \hat{ M } = \hat{\sigma}^2 \cdot ( \hat{h}^2 \Phi + (1-\hat{h}^2) I )$. In the second step, for each $g_i$ and corresponding $X_i = [ 1 | L | g_i ]$, the estimates of the effects and the variance-covariance matrix are respectively obtained as
\begin{equation}
    \hat{\beta}_{i} = (X_i^T \hat{ M }^{-1} X_i)^{-1} X_i^T \hat{ M }^{-1} y,
\label{eq.:GLS}
\end{equation}
and  
$$ Var(\hat{\beta}_i) = \hat{\sigma}^2 \cdot (X_i^T \hat{ M }^{-1} X_i)^{-1},$$
with $i = 1, \ldots, m$, and $m$ is the number of genetic markers considered.

In this work, we consider an extended formulation of this problem to the case of multiple phenotypes, that is, Y is a collection of $t$ vectors, with $y_j$ ($j = 1, \ldots, t$) being a vector corresponding to a specific trait. In this case, trait-specific estimates ${\hat{\sigma}^2_j,\hat{h}^2_j}$ need to be obtained, resulting in $t$ different $\hat{M_j}$s. As the result of the analysis, $m \times t$ vectors of estimates of $\hat{\beta}_{ij}$ and corresponding $Var(\hat{\beta}_{ij})$ are generated. In summary, the problem we are facing is (see Fig.~1 of the main manuscript text for a visual description)
\begin{equation}
\label{eq:probDef}
\left\{
{\begin{aligned}
    \hat{\beta}_{ij} & = (X_i^T \hat{M}_j^{-1} X_i)^{-1} X_i^T \hat{M}_j^{-1} y_j \\
           Var(\hat{\beta}_{ij}) & = \hat{\sigma}^2_j \dot (X_i^T \hat{M}_j^{-1} X_i)^{-1}  \\[2mm]
    \hat{M}_j    & = \hat{\sigma}^2_j (\hat{h}^2_j \Phi + (1 - \hat{h}^2_j) I)
\end{aligned}}
\right.
\;\;\;
{\begin{aligned}
\text{ with } & 1 \le i \le m \\
\text{ and } & 1 \le j \le t.
\end{aligned}}
\end{equation}

\section{Single-instance Algorithms}
In this section we provide an overview of a simplified version of \chol{} and \eig{} to solve a single GLS; the versions tailored for sequences of GLS's are then introduced in Section~\ref{sec:seqs}. In the following, we use $b$ to indicate $\hat \beta$; in both our algorithms, the computation of $Var(b)$ represents an intermediate result towards $b$.  

\paragraph{\chol{}} The approach for \chol{} (Alg.~\ref{alg:single-GLS-chol}) is to first reduce the initial GLS
$b := (X^T \hat{M}^{-1} X)^{-1} X^T \hat{M}^{-1} y$
to a linear least-squares problem, and then solve this via normal equations. Specifically, the algorithm starts by forming
$ \hat{ M } = \hat{\sigma^2} \cdot ( \hat{h^2} \Phi + (1-\hat{h^2}) I )$,
which is known to be symmetric positive definite, and by computing its Cholesky factor $L$.
This leads to the expression
$b := (X^T L^{-T} L^{-1} X)^{-1} X^T L^{-T} L^{-1} y$,
in which two triangular linear systems can be identified and solved
--- $X' := L^{-1} X$ and $y' := L^{-1} y$ ---
thus completing the reduction to the standard least-squares problem
$b := (X'^T X')^{-1} X'^T y'$.
Numerical considerations allow us to safely rely on the Cholesky factorization of
$S := X'^T X'$ without incurring instabilities. The algorithm completes by computing
$b' := X'^T y'$ and solving the linear system $b := S^{-1} b'$, for a total cost of
$\frac{1}{3} n^3 + O( n^2 p )$ for a single GLS.

\begin{center}
\renewcommand{\lstlistingname}{Algorithm}
\begin{minipage}{0.50\linewidth}
\begin{lstlisting}[caption=\chol{}'s approach for a single GLS problem, label=alg:single-GLS-chol, escapechar=!]
   $\hat{ M } := \hat{\sigma^2} \cdot ( \hat{h^2} \Phi + (1-\hat{h^2}) I )$
   $L L^T = \hat{ M }$
   $X := L^{-1} X$
   $y := L^{-1} y$
   $S := X^T X$
   $b := X^T y$
   $b := S^{-1} b$
\end{lstlisting}
\end{minipage}
\end{center}

\paragraph{\eig{}} Instead of forming the matrix $\hat{M}$, Algorithm~\eig{} (Alg.~\ref{alg:single-GLS-eig}) operates on the matrix $\Phi$: At first, it diagonalizes $\Phi$ as $Z W Z^T$, leading to the expression
$$\hat{M} := \hat{\sigma^2} ( \hat{h^2} Z W Z^T + (1 - h^2) I ),$$
with diagonal $W$.
By orthogonality of $Z$, the inverse of $\hat{M}$ can be represented as
$$\hat{M}^{-1} := Z ( \hat{\sigma^2} (\hat{h^2} W + (1 - h^2) I) )^{-1} Z^T$$
and easily computed via
$$D := (\hat{\sigma^2} (\hat{h^2} W + (1 - h^2) I) )^{-1};$$
the solution to the GLS is thus given by
\begin{equation}
b := (X^T Z D^{-1} Z^T X)^{-1} X^T Z D^{-1} Z^T y.
    \label{eq:Eig0}
\end{equation}
Moreover, since $D$ is symmetric positive definite, Eq.~\ref{eq:Eig0} can be rewritten as
$$b :=(X^T Z K K^T Z^T X)^{-1} X^T Z K K^T Z^T y,$$
and the algorithm proceeds by computing $V := X^T Z K$ (matrix-matrix multiplication and scaling), obtaining $b:=(V V^T)^{-1} V K^T Z^T y$. Similar to \chol{}, $b$ is finally obtained through matrix-vector multiplications and a linear system, for a total cost of
$\frac{10}{3} n^3 + O( n^2 p )$.

\begin{center}
\renewcommand{\lstlistingname}{Algorithm}
\begin{minipage}{0.50\linewidth}
\begin{lstlisting}[caption=CLAK-Eig's approach for a single GLS problem, label=alg:single-GLS-eig, escapechar=!]
   $Z W Z^T = \Phi$
   $D := (\hat{\sigma^2} (\hat{h^2} W + (1 - h^2) I) )^{-1}$
   $K K^T = D$
   $V := X^T Z K$
   $S := V V^T$
   $b := V K^T Z^T y$
   $b := S^{-1} b$
\end{lstlisting}
\end{minipage}
\end{center}

\section{Sequences of GLS'} \label{sec:seqs}
As shown in Table\ref{tab:cost}, the strength of \chol{} and \eig{} becomes apparent in the context of 1D and 2D sequences of GLS', corresponding to GWAS with single and multiple phenotypes, respectively. The straightforward approach, which is the only alternative provided by current general-purpose numerical libraries, lies in a loop that utilizes the best performing algorithm for one GLS. No matter how optimized the GLS solver, such an approach is prohibitively expensive for either single or multiple phenotypes, due to the unmanageable complexity of $O( m n^3 )$ and $O( t m n^3 )$, respectively. By contrast, the versions of \chol{} and \eig{} shown in Fig.~\ref{Fig[ALG-seq]} are the product of a number of optimizations aimed at saving intermediate results across successive problems, thus avoiding redundant calculations. For instance, the matrix $X$ is logically split as $[X_L | X_R]$, where $X_L$ includes the intercept and the covariates ($[ 1 | L ]$), while $X_R$ is the collection of all the genetic markers $g_i$. Thanks to these savings, both algorithms achieve a lower overall complexity (Table 1).

\begin{center}
\renewcommand{\lstlistingname}{Algorithm}
\begin{minipage}{0.47\linewidth}
\begin{lstlisting}[caption=CLAK-Chol single-trait analysis, escapechar=!,label=alg:inc-gwas]
   $\hat{ M } := \hat{\sigma^2} \cdot ( \hat{h^2} \Phi + (1-\hat{h^2}) I )$
   $L L^T = \hat{M}$
   $X_L := L^{-1} X_L$
   $X_R := L^{-1} X_R$
   $y := L^{-1} y$
   $S_{TL} := X_L^T X_L$
   $b_T := X_L^T y$
   for i = 1:m
      $S := \left( \renewcommand{\arraystretch}{1.2} \begin{array}{c | c} S_{TL} & \star \\\hline X_{R_i}^T X_L & X_{R_i}^T X_{R_i} \end{array} \right)$
      $b_i := \left( \renewcommand{\arraystretch}{1.2} \begin{array}{c} b_{T} \\\hline X_{R_i}^T y \end{array} \right)$
      $b_i := S_i^{-1} b_i$ !\vspace*{24mm}!
\end{lstlisting}
\end{minipage}
\hfill
\renewcommand{\lstlistingname}{Algorithm}
\begin{minipage}{0.47\linewidth}
\begin{lstlisting}[caption=CLAK-Eig for multi-trait analysis, escapechar=!,label=alg5]
   $Z W Z^T = \Phi$
   $X_L := Z^T X_L$                    
   $X_R := Z^T X_R$                   
   $Y := Z^T Y$                         
   for j = 1:t
      $D := (\hat{\sigma_j}^2 (\hat{h_j}^2 W + (1 - \hat{h_j}^2) I) )^{-1}$
      $K K^T = D$
      $Y_j := K^T Y_j$
      $W_L := K^T X_L$
      $S_{TL} := W_L^T W_L$
      $b_T := W_L^T Y_j$
      for i = 1:m
         $W_R := K^T X_{R_i}$
         $S := \left( \renewcommand{\arraystretch}{1.2} \begin{array}{c | c} S_{TL} & \star \\\hline W_R^T W_L & W_R^T W_R \end{array} \right)$
         $b_{ij} := \left( \renewcommand{\arraystretch}{1.2} \begin{array}{c} b_{T} \\\hline W_R^T Y_j \end{array} \right)$
         $b_{ij} := S^{-1} b_{ij}$
\end{lstlisting}
\end{minipage}
\end{center}

Unfortunately, the reduced complexity of algorithms \chol{} and \eig{} is not enough to guarantee high-performance implementations. It is well known that in terms of execution time, a straightforward translation of algorithms into code and a carefully assembled routine often differ by at least one order of magnitude. In other words, the benefits inherent to our new algorithms might go unnoticed unless paired with state-of-the-art implementation techniques. In this section we detail our strategy to attain high-performance routines.

Both \chol{} and \eig{} are entirely expressed in terms of standard linear algebra operations such as matrix products and matrix factorizations, provided by the BLAS~\cite{BLAS3} and LAPACK~\cite{laug} libraries. Since LAPACK itself is built in terms of BLAS kernels, these are the main responsible for the overall performance of an algorithm. BLAS consists of a relatively small set of highly optimized kernels, organized in three levels, corresponding to vector, vector-matrix, and matrix-matrix operations, respectively. A common misconception is that all the BLAS kernels, across the three levels, attain a comparable (and high) level of efficiency.
Instead, it is only BLAS-3 --when operating on large matrices-- that fully exploits the processors' potential; as an example, the matrix-vector multiplication, matrix-matrix multiplication on small matrices, and matrix-matrix-multiplication on large matrices attain an efficiency of 10\%, $\approx 40\%$, and more than 95\%, respectively. In this context, the linear systems $X := L^{-1} X$ in \chol{} (Line 3 of Algorithm 1), to be solved for each individual SNP, should ideally be aggregated into a single ---very large--- linear system $X_R := L^{-1} X_R$, in which $X_R$ is the collection of the genetic markers of all SNPs (Line 3 of Algorithm 3). An analogous comment is valid for \eig{} (see the multiplications at Lines 3 and 4 of Algorithm 4), in which the number of markers accessed at once is a user-configurable input parameter.

Since all the current computing platforms include multicore processors, we now briefly discuss how to take advantage of this architecture. An effective practice is to invoke a multi-threaded version of the BLAS library. Both in Algorithm 3 and 4, we rely on such a solution for the sections leading up to the outer loop (Lines 1--6 and 1--4, respectively). In the remainder of the algorithms (Lines 7--11 and 5--13), due to the lack of matrix-matrix operations, we instead apply a thread-based parallelization in conjunction with single-threaded BLAS. This mixed use of multi-threading becomes more and more effective as the number of available computing cores increases, leading to speedups up to 10 or even 20\%.

\subsection{Space requirement}

\paragraph{\chol{}}
To form and factor the variance matrix, the algorithm uses $n^2$ memory (Lines 1--2 in Algorithm 3). The overall space requirement is determined by the triangular solve (Line 4), which necessitates the full $L$ and a portion of $X$ to reside in memory at the same time. This operation is performed in a streaming fashion ---operating on $k$ SNPs at the time--- and overwriting $X$. All the other instructions do not require any extra space. In total, \chol{} uses about $n^2 + k n p$ memory.

\paragraph{\eig{}}
The initial eigendecomposition (Line 1 in Algorithm 4) needs $2 n^2$ memory. The following matrix-matrix multiplications (Lines 2, 3 and 8) overwrite the input matrices and again are performed in a streaming fashion; in terms of space, the analysis is similar to that for the triangular solve in \chol{}. The remaining instructions do not affect the overall memory usage. In total, the space requirement is $2 n^2 + k n p$.   

If memory is a limiting factor, one can set $k$ to 1 in either algorithm, possibly at the cost of exposed data movement. Moreover, by using a different eigensolver, the eigenvectors $Z$ can overwrite the input matrix $\Phi$, effectively saving $n^2$ memory. These considerations indicate that our algorithms are capable of solving GWAS of any size, as long as the kinship matrix and one SNP fit in RAM.

\subsection{Time complexity}

As it was described in Sec.~\ref{MixModels}, before the sequence of GLS can be solved, one needs to estimate the parameters related to the random part of the model, namely the heritability ($h^2$) and the total variance ($\sigma^2$). Software packages such as GenABEL~\cite{genabel}, EMMAX, and FaST-LMM use algorithms similar to our \eig{}. First, the eigendecomposition of the kinship matrix is performed, for a computational cost of $O(n^3)$. Then, the model parameters are estimated using an iterative procedure based on the Maximum Likelihood (GenABEL, FaST-LMM) or Restricted Maximum Likelihood (EMMAX) methods, for a cost of $v n p^2$, where $v$ is the average number of iterations required to reach convergence. In total, the parameter estimation has a complexity of $\frac{10}{3} n^3 + O(v n p^2)$ operations.

By contrast, when our \chol{} algorithm is used, the initial estimation requires a Cholesky factorization of the kinship matrix at each of the $v$ iterations, for a total cost of $\frac{1}{3} v n^3$ operations. Due to the reduced cost and better performance of such a factorization, this strategy is advantageous when $v < \approx 30$, a condition commonly verified in practice.

In the second step, a 1D or 2D sequence of GLS is solved, corresponding to single and multiple phenotype analysis. For 1D sequences, all the considered methods share the same asymptotic time complexity, but the constant factor for \chol{} is the lowest, yielding at least a 4-fold speedup. In 2D sequences, EMMAX, FaST-LMM and GenABEL simply tackle each individual trait independently, one after another, for a total complexity of $t m p n^2$. By exploiting the common structure of the variance-covariance matrix of different phenotypes,
\eig{} reduces the complexity by a factor of $n$, down to $t m p n$. As a result, our algorithm outperforms the other methods by a factor of 100 and more.

\begin{table*}[!h]
\centering
\setlength\extrarowheight{2pt}
\renewcommand{\arraystretch}{1.3}
\begin{tabular}{| c | c | c | c |} \toprule
& Estimation of $\sigma^2$ and $h^2$ & 1D Sequence   & 2D Sequence \\\midrule
 Naive       & --- & $O(m n^3)$   & $O( m t n^3)$ \\
 \chol{}   & {\sc chol-based}   & $\mathbf{O(m p n^2)}$ & --- \\
 \eig{}    & {\sc eigen-based}  & --- & $\mathbf{O(t m p n)}$\\
 FaST-LMM    & {\sc eigen-based}  & $O(m p n^2)$ & $O(t m p n^2)$\\
 GWFGLS      & {\sc eigen-based}  & $O(m p n^2)$ & $O(t m p n^2)$ \\
 EMMAX       & {\sc eigen-based}  & $O(m p n^2)$ & $O(t m p n^2)$ \\\bottomrule
\end{tabular}
\caption{
Computational costs for single GLS, as well as 1-dimensional and 2-dimensional sequences of GLSs. The variables $n, m$ and $t$ denote the sample size, the number of genetic markers, and the number of traits, respectively. $p$ is the width of the $X$ matrix, which is determined by the number of fixed covariates used (e.g. age and sex) and the genetic model assumed.
}
\label{tab:cost}
\end{table*}

\section{Computing environment}

All the computing tests were run on a SMP system consisting of two Intel Xeon X5675 6-core processors, operating at a frequency of 3.06 GHz. The system is equipped with 96GB of RAM
and 1TB of disk as secondary memory.\footnote{Recall that our algorithms do not require large amounts of available memory: As long as the kinship matrix fits in memory, they will complete.} The routines were compiled with the GNU C Compiler (gcc v4.4.5) and linked to Intel's MKL multi-threaded library (v10.3). For double-buffering, our out-of-core routines make use of the AIO library, one of the standard libraries on UNIX systems. The available multi-core parallelism is exploited through MKL's multi-threaded BLAS and OpenMP's pragma directives.

\subsection{Simulated data}

A data set for GWAS can be characterized by the number of individuals in the sample ($n$), the number of measured and/or imputed SNPs ($m$) to be tested, the number of outcomes to be analyzed ($t$), and the number of covariates ($p$) to be included in the model. In current GWAS, a typical scenario consists of a few covariates (for example, two, such as sex and age), $10^5--10^7$ SNPs, and thousands or tens of thousands individuals; also, only one or a limited number of outcomes are studied.
In our experiments, we assumed that the number of covariates is two (p=2), and we varied the three other characteristics ($n, m, t$) of the data set leading to these three scenarios.
\begin{enumerate}[(A)]
  \item The number of SNPs was fixed to $m = 10{,}000{,}000$ and one single outcome ($t = 1$) was studied. As sample size, we used $1{,}000$, $5{,}000$, $10{,}000$, $20{,}000$, and $40{,}000$. The latter test represents a scenario with large number of individuals.
  \item The number of individuals was fixed to $n = 10,000$ and one single outcome ($t=1$) was studied. The number of markers $m$ to be analyzed was set to $1{,}000{,}000$, $10{,}000{,}000$, and $36{,}000{,}000$. The latter test is a scenario that represents a whole genome resequencing.
  \item The number of individuals and markers were fixed to $n=1{,}000$ and $m=1{,}000{,}000$, respectively. The number of outcomes $t$ studied varied ($1$, $10$, $100$, $1{,}000$, $10{,}000$, and $100{,}000$), corresponding to an Omics analysis.
\end{enumerate}

For testing purposes, we generated artificial data sets which met pre-specified values of $t$, $m$, $p$, and $n$.
